\documentclass{article} 
\usepackage{color}
\usepackage{amsmath,amsfonts,amsthm,amssymb}
\usepackage{graphicx}
\usepackage{fullpage}

\definecolor{webgreen}{rgb}{0,.5,0}
\definecolor{webblue}{rgb}{0,0,.5}

\DeclareMathOperator{\Exp}{Exp}
\DeclareMathOperator{\Var}{Var}

\numberwithin{equation}{section}

\newtheorem{theorem}{Theorem}
\newtheorem{lemma}[theorem]{Lemma}

\newtheorem{definition}{Definition}
\DeclareMathAlphabet{\varmathbb}{U}{bbold}{m}{n}
\newcommand{\one}{\varmathbb 1}
\renewcommand{\vec}[1]{\mathbf{#1}}

\newcommand{\remove}[1]{}

\newcommand{\C}{\mathbb{C}}

\newcommand{\Z}{\mathbb{Z}}

\newcommand{\ket}[1]{\left| #1 \right\rangle}
\newcommand{\bra}[1]{\left\langle #1 \right|}

\newcommand{\CG}{\C[G]}
\newcommand{\cg}{\frac{1}{\sqrt{|G|}}}
\newcommand{\tr}{\textbf{tr}\,}
\newcommand{\im}{\textbf{im}\;}
\newcommand{\rank}{\textbf{rk}\;}
\newcommand{\norm}[1]{\left\| #1 \right\|}
\newcommand{\abs}[1]{\left| #1 \right|}
\newcommand{\U}{\textsf{U}}
\newcommand{\GL}{\textsf{GL}}

\newcommand{\supp}{{\rm supp}}
\newcommand{\poly}{{\rm poly}}
\newcommand{\chrangle}[1]{\rangle_{#1}}
\newcommand{\chlangle}{\langle}

\newcommand{\vb}{\vec{b}}
\newcommand{\ve}{\vec{e}}

\newcommand{\vu}{\vec{u}}

\newcommand{\vv}{\vec{v}}
\newcommand{\bad}{B_{\textrm{bad}}}

\newcommand{\hook}{\text{hook}}

\newcommand{\Ind}{\text{Ind}}
\title{The Symmetric Group Defies Strong Fourier Sampling: Part I}

\author{Cristopher Moore \\
  \textsf{moore@cs.unm.edu}\\
  Department of Computer Science\\
  University of New Mexico
  \and
  Alexander Russell\\
  \textsf{acr@cse.uconn.edu}\\
  Department of Computer Science and Engineering\\
  University of Connecticut
  \and
  Leonard J. Schulman\\
  \textsf{schulman@cs.caltech.edu}\\
  Computer Science Department\\
  California Institute of Technology}

\begin{document}
\maketitle 

\begin{abstract}
  We resolve the question of whether Fourier sampling can efficiently
  solve the hidden subgroup problem. Specifically, we show that the
  hidden subgroup problem over the symmetric group cannot be
  efficiently solved by strong Fourier sampling, even if one may
  perform an arbitrary POVM on the coset state.
  These results apply to the special case relevant to the Graph
  Isomorphism problem.  
\end{abstract}

\section{Introduction: the hidden subgroup problem}

Many problems of interest in quantum computing can be reduced to an
instance of the \emph{Hidden Subgroup Problem} (HSP).  We are given a
group $G$ and a function $f$ with the promise that, for some subgroup
$H \subseteq G$, $f$ is invariant precisely under translation by $H$: that
is, $f$ is constant on the cosets of $H$ and takes distinct values on
distinct cosets.  We then wish to determine the subgroup $H$ by
querying $f$.

For example, in Simon's problem~\cite{Simon97}, $G=\Z_2^n$ and $f$ is
an oracle such that, for some $y$, $f(x) = f(x+y)$ for all $x$; in
this case $H=\{0,y\}$ and we wish to identify $y$.  In Shor's
factoring algorithm~\cite{Shor97} $G$ is the group $\Z_n^*$ where $n$
is the number we wish to factor, $f(x) = r^x \bmod n$ for a random $r
< n$, and $H$ is the subgroup of $\Z_n^*$ whose index is the
multiplicative order of $r$.
Both Simon's and Shor's algorithms use the following approach,
referred to as the \emph{standard method} or \emph{Fourier sampling}
\cite{BernsteinV93}.

\begin{description}
\item[Step 1.] Prepare two registers, the first in a uniform
  superposition over the elements of $G$ and the second with the value
  zero, yielding the state
  $$
  \psi_1 = \cg \sum_{g \in G} \ket{g} \otimes \ket{0} \enspace.
$$
\item[Step 2.] Query (or calculate) the function $f$ defined on $G$
  and XOR it with the second register.  This entangles the two
  registers and results in the state
  $$
  \psi_2 = \cg \sum_{g \in G} \ket{g} \otimes \ket{f(g)} \enspace.
$$
\item[Step 3.] Measure the second register.  This puts the first
  register in a uniform superposition over one of $f$'s level sets,
  i.e., one of the cosets of $H$, and disentangles it from the second
  register.  If we observe the value $f(c)$, we have the state $\psi_3 \otimes
  \ket{f(c)}$ where
$$
\psi_3 = \ket{cH}
= \frac{1}{\sqrt{|H|}} \; \sum_{h \in H} \ket{ch}
 \enspace.
 $$
 Alternately, we can view the first register as being in a mixed
 state with density matrix
\[ \frac{1}{|G|} \sum_{c} \ket{cH} \bra{cH} \]
where the sum includes one representative $c$ for each of $H$'s
cosets.
  
\item[Step 4.] Carry out the quantum Fourier transform on $\psi_3$ and
  measure the result.
\end{description}
(Note that in Shor's algorithm, since $|\Z_n^*|$ is unknown, the
Fourier transform is performed over $\Z_q$ for some $q=\poly(n)$; see
\cite{Shor97} or \cite{HalesH99,HalesH00}.)

In both Simon's and Shor's algorithms, the group $G$ is abelian; it is
not hard to see that, in this abelian case, a polynomial number (i.e.,
polynomial in $\log |G|$) of experiments of this type determine $H$.
In essence, each experiment yields a random element of the dual space
$H^\perp$ perpendicular to $H$'s characteristic function, and as soon as
these elements span $H^\perp$ we are done.

While the \emph{nonabelian} hidden subgroup problem appears to be much
more difficult, it has very attractive applications.  In particular,
solving the HSP for the symmetric group $S_n$ would provide an
efficient quantum algorithm for the Graph Automorphism and Graph
Isomorphism problems (see e.g.\ Jozsa~\cite{Jozsa00} for a review).
Another important motivation is the relationship between the HSP over
the dihedral group with hidden shift problems~\cite{vanDamHI03} and
cryptographically important cases of the Shortest Lattice Vector
problem~\cite{Regev}.

So far, algorithms for the HSP are only known for a few families of
nonabelian groups, including wreath products $\Z_2^k\; \wr\;
\Z_2$~\cite{RoettelerB98}; more generally, semidirect products $K
\ltimes \Z_2^k$ where $K$ is of polynomial size, and groups whose
commutator subgroup is of polynomial size~\cite{IvanyosMS01};
``smoothly solvable'' groups~\cite{FriedlIMSS02}; and some semidirect
products of cyclic groups~\cite{InuiLeGall}.  Ettinger and
H{\o}yer~\cite{EttingerH98} provided another type of result, by showing
that Fourier sampling can solve the HSP for the dihedral groups $D_n$
in an \emph{information-theoretic} sense.  That is, a polynomial
number of experiments gives enough information to reconstruct the
subgroup, though it is unfortunately unknown how to determine $H$ from
this information in polynomial time.

To discuss Fourier sampling for a nonabelian group $G$, one needs to
develop the Fourier transform over $G$ which relies on the group's
\emph{linear representations}. For abelian groups, the Fourier basis
functions are homomorphisms $\phi:G \to \C$ such as the familiar
exponential function $\phi_k(x) = e^{2\pi i kx/n}$ for the cyclic group
$\Z_n$.  In the nonabelian case, there are not enough such
homomorphisms to span the space of all $\C$-valued functions on $G$;
to complete the picture, one introduces \emph{representations} of the
group, namely homomorphisms $\rho:G \to \U(V)$ where $\U(V)$ is the group
of unitary matrices acting on some $\C$-vector space $V$ of dimension
$d_\rho$. It suffices to consider \emph{irreducible} representations,
namely those for which no nontrivial subspace of $V$ is fixed by the
various operators $\rho(g)$.
Once a basis for each irreducible $\rho$ is chosen, the matrix elements
$\rho_{ij}$ provide an orthogonal basis for the vector space of all
$\C$-valued functions on $G$.

\begin{sloppypar}
  The quantum Fourier transform then consists of transforming
  (unit-length) vectors in $\CG = \{ \sum_{g \in G} \alpha_g \ket{g} \mid \alpha_g
  \in \C\}$ from the basis $\{ \ket{g} \mid g \in G \}$ to the basis $\{
  \ket{\rho,i,j} \}$ where $\rho$ is the name of an irreducible
  representation and $1 \leq i, j \leq d_\rho$ index a row and column (in a
  chosen basis for $V$).  Indeed, this transformation can be carried
  out efficiently for a wide variety of
  groups~\cite{Beals97,Hoyer97,MooreRR04}. Note, however, that a
  nonabelian group $G$ does not distinguish any specific basis for its
  irreducible representations which necessitates a rather dramatic
  choice on the part of the transform designer. Indeed, careful basis
  selection appears to be critical for obtaining efficient Fourier
  transforms for the groups mentioned above.
\end{sloppypar}

Perhaps the most fundamental question concerning the hidden subgroup
problem is whether there is always a basis for the representations of
$G$ such that measuring in this basis (in Step 4, above) 
provides enough information to determine the subgroup $H$. This
framework is known as \emph{strong Fourier sampling}. In this article,
we answer this question in the negative, showing that natural
subgroups of $S_n$ cannot be determined by this process; in fact, 
we show this for an even more general model, where we perform 
an arbitrary positive operator-valued measurement (POVM) on coset states 
$\ket{cH}$.  We emphasize
that the subgroups on which we focus are among the most important
special cases of the HSP, as they are those to which Graph Isomorphism
naturally reduces.

\paragraph{Related work.} The terminology ``strong Fourier sampling''~\cite{GrigniSVV01} 
was invented to distinguish this approach from the natural variant,
called \emph{weak Fourier sampling}, where one only measures the name
of the representation $\rho$, and ignores the row and column
information.  Weak Fourier sampling is basis-independent, making it
attractive from the standpoint of analysis; however, it cannot
distinguish conjugate subgroups from each other, and Hallgren, Russell
and Ta-Shma~\cite{HallgrenRT00} showed that it cannot distinguish the
trivial subgroup from an order-2 subgroup consisting of $n/2$ disjoint
transpositions.  Specifically, they used character bounds to show that
the probability distribution obtained on representation names for the
trivial and order-2 subgroups are exponentially close in total
variation distance: it requires an exponential number of such
experiments to distinguish them.  Kempe and Shalev~\cite{KempeS} have
generalized this result to other conjugacy classes, and conjecture
that one can do no better than classical computation with this
approach.

In an effort to shed light on the power of strong Fourier sampling,
Grigni, Schulman, Vazirani and Vazirani~\cite{GrigniSVV01} showed
that, for groups such as $S_n$, measuring in a \emph{random} basis
yields an exponentially small amount of information.  This can be
explained, roughly, by the fact that projecting a vector into a
sufficiently high-dimensional random subspace results in tightly
concentrated length.  On the other hand, Moore, Rockmore, Russell and
Schulman~\cite{MooreRRS04} showed that for the affine and $q$-hedral
groups, measuring in a well-chosen basis can solve the HSP (at least
information-theoretically) in cases where random bases cannot.

\paragraph{Our contribution.} 
In this paper we show that strong Fourier sampling, in an arbitrary basis 
of the algorithm designer's choice, cannot solve the HSP for $S_n$.  
As in~\cite{HallgrenRT00} we focus on
order-2 subgroups of the form $\{1,m\}$ where $m$ is an involution consisting of
$n/2$ disjoint transpositions; we remark that if we fix two rigid
connected graphs of size $n/2$ and consider permutations of their
disjoint union, then the hidden subgroup is of this form if the graphs
are isomorphic and trivial if they are not. Then we show that 
strong Fourier sampling---and more generally, 
arbitrary measurements of coset states---cannot distinguish
these subgroups from each other, or from the trivial subgroup, without
an exponential number of experiments.

We remark that our results do not preclude the existence of an efficient
quantum algorithm for the HSP on $S_n$.  Rather, they force us to consider
\emph{multi-register} algorithms, in which we prepare multiple coset
states and subject them to entangled measurements.  Ettinger, H{\o}yer
and Knill~\cite{EttingerHK04} showed that the HSP on arbitrary groups
can be solved information-theoretically with a polynomial number of
registers, and the authors have shown how to carry out such a measurement 
in the Fourier basis~\cite{MooreR:banff}.  Kuperberg~\cite{Kuperberg03} devised a
subexponential ($2^{O(\sqrt{\log n})}$) algorithm for the HSP on the
dihedral group $D_n$ that works by performing entangled measurements on
two registers at a time, Bacon, Childs, and van
Dam~\cite{BaconCvD} have determined the optimal multiregister
measurement for the dihedral group, and the present authors have generalized 
this to all cases where the $H$ and $G$ form a Gel'fand pair~\cite{MooreR:gelfand}.

Whether a similar approach can be taken to the symmetric group is a
major open question.  In a companion paper~\cite{MooreR:part2}, the first two authors 
take a step towards answering this question by showing that if we perform
arbitrary entangled measurements over pairs of registers,
distinguishing $H=\{1,m\}$ from the trivial group in $S_n$ requires a
superpolynomial number (specifically, $e^{\Omega(\sqrt{n}/\log n)}$) of
experiments.

\section{Fourier analysis over finite groups}
\label{sec:representation-theory}

We briefly discuss the elements of the representation theory of finite
groups. Our treatment is primarily for the purposes of setting down
notation; we refer the reader to~\cite{FultonH91,Serre77} for complete
accounts.

Let $G$ be a finite group. A \emph{representation} $\rho$ of $G$ is a
homomorphism $\rho: G \to \U(V)$, where $V$ is a finite dimensional
Hilbert space and $\U(V)$ is the group of unitary operators on $V$.
The \emph{dimension} of $\rho$, denoted $d_\rho$, is the dimension of the
vector space $V$. By choosing a basis for $V$, then, each $\rho(g)$ is
associated with a unitary matrix $[\rho(g)]$ so that for every $g, h \in
G$, $[\rho(gh)] = [\rho(g)] \cdot [\rho(h)]$ where $\cdot$ denotes matrix
multiplication.

Fixing a representation $\rho: G \to \U(V)$, we say that a subspace $W \subset
V$ is \emph{(left)-invariant} if $\rho(g) W \subset W$ for all $g \in G$; observe
that in this case the restriction $\rho_W: G \to \U(W)$, given by
restricting each $\rho(g)$ to $W$, is also a representation. When $\rho$
has no invariant subspaces other than the trivial space $\{ \vec{0}
\}$ and $V$, $\rho$ is said to be \emph{irreducible}. When $\rho$ is
irreducible, \emph{Schur's lemma} asserts that the centralizer of the
subgroup $\im \rho \subset \U(V) \subset \GL(V)$---that is, the set of $A \in
\GL(V)$ such that $A\rho(g) = \rho(g)A$ for all $g \in G$---consists only
of the scalar matrices $\{ c\one \mid c \in \C \}$. We use this fact below.

In the case when $\rho$ is \emph{not}
irreducible, then, there is a nontrivial invariant subspace $W \subset V$
and, as the inner product $\langle \cdot, \cdot\rangle$ is invariant under the 
unitary maps $\rho(g)$, it is immediate that the dual subspace
$$
W^\perp = \{ \vec{u} \mid \forall \vec{w} \in W, \langle\vec{u},\vec{w}\rangle = 0\}
$$
is also invariant.  Associated with the decomposition $V = W \oplus
W^\perp$ is the natural decomposition of the operators $\rho(g) = \rho_W(g)
\oplus \rho_{W^\perp}(g)$. By repeating this process, any representation $\rho: G
\to \U(V)$ may be decomposed into a direct sum of irreducible
representations: we write $\rho = \sigma_1 \oplus \cdots \oplus \sigma_k$ and, for the
$\sigma_i$ appearing at least once in this decomposition, $\sigma_i \prec \rho$.
In general, a given $\sigma$ can appear multiply in this decomposition,
in the sense that $\rho$ may have an invariant subspace isomorphic to
the direct sum of $a_\sigma$ copies of $\sigma$.  In this case $a_\sigma$ is
called the \emph{multiplicity} of $\sigma_i$ in $\rho$, and we write $\rho =
\bigoplus_{\sigma \prec \rho} a_\sigma \sigma$ where $a_\sigma \sigma = \underbrace{\sigma \oplus \cdots \oplus
  \sigma}_{a_\sigma}$.

If two representations $\rho$ and $\sigma$ are the same up to a unitary
change of basis, we say that they are \emph{equivalent}. It is a fact
that any finite group $G$ has a finite number of distinct irreducible
representations up to equivalence and, for a group $G$, we let
$\widehat{G}$ denote a set of representations containing exactly one
from each equivalence class. The irreducible representations of $G$
give rise to the Fourier transform. Specifically, for a function $f: G
\to \C$ and an element $\rho \in \widehat{G}$, define the \emph{Fourier
  transform of $f$ at $\rho$} to be
$$
\hat{f}(\rho) = \sqrt{\frac{d_\rho}{|G|}} \sum_{g \in G} f(g)\rho(g)\enspace.
$$
The leading coefficients are chosen to make the transform unitary,
so that it preserves inner products:
$$
\langle f_1, f_2\rangle = \sum_g f_1^*(g)f_2(g) 
= \sum_{\rho \in \widehat{G}}\tr \!\left(\hat{f_1}(\rho)^\dagger \cdot \hat{f_2}(\rho)\right)\enspace.
$$

There is a natural product operation on representations: if $\rho: G \to
\U(V)$ and $\sigma: G \to \U(W)$ are representations of $G$, we may define
a new representation $\rho \otimes \sigma: G \to \U(V \otimes W)$ by extending the rule
$\rho \otimes \sigma(g): \vec{u} \otimes \vec{v} \mapsto \rho(g)\vec{u} \otimes \sigma(g)\vec{v}$.  In
general, the representation $\rho \otimes \sigma$ is not irreducible, even when
both $\rho$ and $\sigma$ are. This leads to the \emph{Clebsch-Gordan
  problem}, that of decomposing $\rho \otimes \sigma$ into irreducibles.

For a representation $\rho$ we define the \emph{character} of $\rho$,
denoted $\chi_\rho$, to be the function $\chi_\rho: G \to \C$ given by
$\chi_\rho(g) = \tr \rho(g)$. As the trace of a linear operator is invariant
under conjugation, characters are constant on the conjugacy classes of
$G$; for a conjugacy class $A = \{ gag^{-1} \mid g \in G\}$, we define
$\chi(A) = \chi(a)$. Characters are a powerful tool for reasoning about
the decomposition of reducible representations. In particular, when
$\rho = \bigoplus_i \sigma_i$ we have $\chi_\rho = \sum_i \chi_{\sigma_i}$ and, moreover, for
$\rho, \sigma \in \widehat{G}$, we have the orthogonality conditions
$$
\chlangle \chi_\rho, \chi_\sigma \chrangle{G} = \frac{1}{|G|} \sum_{g \in G} \chi_\rho(g)\chi_\sigma(g)^*
= \begin{cases} 1 & \rho = \sigma\enspace,\\
  0 & \rho \neq \sigma\enspace. \end{cases}
$$
Then for an irreducible representation $\rho$ and
representation $\sigma$, $\chlangle \chi_\rho, \chi_\sigma \chrangle{G}$ is equal to 
the multiplicity with which $\rho$ appears in the decomposition of
$\sigma$.  For example, since $\chi_{\rho \otimes \sigma}(g) = \chi_\rho(g) \cdot \chi_\sigma(g)$, 
the multiplicity of $\tau$ in $\rho \otimes \sigma$ is $\chlangle \chi_\tau, \chi_\rho \chi_\sigma \chrangle{G}$

In general, we can consider subspaces of $\CG$ that are invariant
under left multiplication, right multiplication, or both; these
subspaces are called \emph{left-}, \emph{right-}, or
\emph{bi-invariant} respectively.  Each $\rho \in \widehat{G}$
corresponds to a $d_\rho^2$-dimensional bi-invariant subspace of $\CG$,
which can be broken up further into $d_\rho$ $d_\rho$-dimensional
left-invariant subspaces, or (transversely) $d_\rho$ $d_\rho$-dimensional
right-invariant subspaces.  However, this decomposition is not unique.
If $\rho$ acts on a vector space $V$, then choosing an orthonormal basis
for $V$ allows us to view $\rho(g)$ as a $d_\rho \times d_\rho$ matrix.  Then
$\rho$ acts on the $d_\rho^2$-dimensional space of such matrices by left
or right multiplication, and the columns and rows correspond to left-
and right-invariant spaces respectively.  More generally, each
left-invariant subspace isomorphic to $\rho$ corresponds to a unit
vector $\vb \in V$.

\section{The structure of the optimal measurement}
\label{sec:optimal}

In this section we show that starting with a random coset state, the
optimal one-register measurement for the hidden subgroup problem is
precisely an instance of strong Fourier sampling (possibly in an 
over-complete basis).  This has been
pointed out several times in the past, at varying levels of
explicitness~\cite{Ip,Kuperberg03}; we state it here for completeness.
Everything we say in this section is true for the hidden subgroup
problem in general.  However, for simplicity we focus on the special
case of the hidden subgroup problem called the \emph{hidden conjugate
  problem} in~\cite{MooreRRS04}: there is a (non-normal) subgroup $H$,
and we are promised that the hidden subgroup is one of its conjugates,
$H^g = g^{-1} H g$ for some $g \in G$.

We may treat the states arising after Step 3 of the procedure above as
elements of the group algebra $\CG$.  We use the notation $\ket{g}
= 1 \cdot g \in \CG$ so that the vectors $\ket{g}$ form an orthonormal
basis for $\CG$.  Given a set $S \subset
G$, $\ket{S}$ denotes a uniform superposition over the elements of
$S$, $\ket{S} = (1/\sqrt{|S|}) \sum_{s \in S} \ket{s}$.

\subsection{The optimal POVM consists of strong Fourier sampling}

The most general type of measurement allowed in quantum mechanics
is a \emph{positive operator-valued measurement}
(POVM).  A POVM with a set of possible outcomes $J$ 
consists of a set of positive operators $\{ M_j \mid j \in J\}$
subject to the completeness condition,
\begin{equation} 
\label{eq:complete}
\sum_j M_j = \one \enspace.
\end{equation}
Since positive operators are self-adjoint, they can be
orthogonally diagonalized, and since their eigenvalues are positive, they
may be written as a positive linear combination of projection
operators (see e.g.~\cite[\S10]{Roman92}).  Any POVM 
may thus be refined so that each $M_j = a_j \mu_j$ 
where $\mu_j$ is a projection operator and $a_j$ is positive and real.

The result of this measurement on the state $\ket{\psi}$ is a random
variable, taking values in $J$, that is equal to $j \in J$ with
probability $P_j = a_j \bra{\psi} \mu_j \ket{\psi}$. Note that outcomes $j$
need not correspond to subgroups directly; the algorithm designer is
free to carry out a polynomial number $t$ of experiments, observing
outcomes $j_1, \ldots, j_t$, and then apply some additional computation to
find the most likely subgroup given these observations.

If $g$ is chosen from $G$ uniformly so that the hidden subgroup is a
uniformly random conjugate of $H$, we wish to find a POVM that
maximizes the probability of correctly identifying $g$ from the coset
state $\ket{H^g}$.  (Of course, to identify a conjugate $H^g$, we only
need to specify $g$ up to an element of the normalizer of $H$.)  Since
a random left coset of $H^g$ can be written $cgH^g = cHg$ for a random
$c \in G$, the probability we observe outcome $j$ is
\begin{equation}
\label{eq:pj}
 P_j = a_j \frac{1}{|G|} \sum_{c \in G} \bra{cHg} \mu_j \ket{cHg}\enspace.
\end{equation}
Ip~\cite{Ip} observed that in the special case that each outcome $j$
corresponds to a subgroup, maximizing the probability that $j$ is
correct subject to the constraint~\eqref{eq:complete} gives a
semi-definite program.  Since such programs are convex, the optimum is
unique and is a fixed point of any symmetries possessed by the
problem.

However, our proof relies on an elementary ``symmetrization''
argument.  Given a group element $x \in G$, let $L_x \ket{g} =
\ket{xg}$ denote the unitary matrix corresponding to left group
multiplication by $x$.  In particular, applying $L_x$ maps one left
coset onto another: $\ket{cHg} = L_c \ket{Hg}$.  Writing
$$
P_j = a_j \frac{1}{|G|} \sum_{c \in G} \bra{cHg} \mu_j \ket{cHg} =
a_j \left\langle Hg \left| \frac{1}{|G|} \sum_{c \in G} L_c^\dagger \mu_j L_c \right| Hg \right\rangle
\enspace ,
$$
we conclude that replacing $\mu_j$ for each $j$ with the
symmetrization
\[ \mu'_j = \frac{1}{|G|} \sum_{g \in G} L_g^\dagger \mu_j L_g \]
does not change the resulting probability distribution $P_j$.  Since
$\mu'_j$ commutes with $L_x$ for every $x \in G$ and provides exactly
the same information as the original $\mu_j$, we may assume without
loss of generality that the optimal POVM commutes with $L_x$ for every
$x \in G$.

It is easy to see that any projection operator that commutes with left
multiplication projects onto a left-invariant subspace of $\CG$, and
we can further refine the POVM so that each $\mu_j$ projects onto an
\emph{irreducible} left-invariant subspace.  Each such space is
contained in the bi-invariant subspace corresponding to some
irreducible representation $\rho$, in which case we write $\im \mu_j \subseteq
\rho$.  As discussed in Section~\ref{sec:representation-theory}, a given
irreducible left-invariant subspace corresponds to some unit vector
$\vb$ in the vector space $V$ on which $\rho$ acts.  Thus we can write
\[ \mu_j = \ket{\vb_j} \bra{\vb_j} \otimes \frac{1}{d_\rho} \one_{d_\rho} \]
where $\one_{d_\rho}$ acts within the left-invariant subspace.  Let
$B=\{ \vb_j \mid \im \mu_j \in \rho \}$; then~\eqref{eq:complete} implies a
completeness condition for each $\rho \in \widehat{G}$,
\begin{equation}
\label{eq:completerho}
 \sum_{\vb_j \in B} a_j \ket{\vb_j} \bra{\vb_j} 
\end{equation}
and so $B$ is a (possibly over-complete) basis for $V$.  In other
words, the optimal POVM consists of first measuring the representation
name $\rho$, and then performing a POVM on the vector space $V$ with
possible outcomes $B$.  Another way to see this is to regard the
choice of coset as a mixed state; then its density matrix is
block-diagonal in the Fourier basis, and so as Kuperberg puts
it~\cite{Kuperberg03} measuring the representation name ``sacrifices
no entropy.''

We note that in the special case that this POVM is a von Neumann
measurement---that is, when the $B$ is an orthonormal basis for
$V$---then it corresponds to measuring the column of $\rho$ in that
basis, which is how strong Fourier sampling is usually defined.  (As
pointed out in~\cite{GrigniSVV01}, nothing is gained by measuring the
row, since we have a random left coset $cHg$ and left-multiplying by a
random element $c$ in an irreducible representation completely mixes
the probability across the rows in each column.  Here this is
reflected by the fact that $\mu_j$ is a scalar in each left-invariant
subspace.)  However, in general the optimal measurement might consist
of an over-complete basis, or \emph{frame}, in each $\rho$, consisting
of the vectors $\{ \vb_j \}$ and the weights $a_j$.

Now that we know $\mu_j$ takes this form, let us change notation.
Given $\rho \in \widehat{G}$ acting on a vector space $V$ and a unit
vector $\vb \in V$, let $\Pi^\rho_{\vb} = \ket{\vb} \bra{\vb} \otimes
\one_{d_\rho}$ denote the projection operator onto the left-invariant
subspace corresponding to $\vb$.  Then $\mu_j = \Pi^\rho_{\vb_j}$,
and~\eqref{eq:pj} becomes
\begin{equation}
\label{eq:pj2} 
 P_j = a_j \frac{1}{|G|} \sum_{c \in G} \norm{\Pi^\rho_{\vb_j} \ket{cHg}}^2 
= a_j \norm{\Pi^\rho_{\vb_j} \ket{Hg}}^2 \enspace . 
\end{equation}
We can write this as the product of the probability $P(\rho)$ that we
observe $\rho$, times the conditional probability $P(\rho,\vb_j)$ that we
observe $\vb_j$.  Note that by~\eqref{eq:completerho},
\[ \Pi^\rho = \sum_{\vb_j \in B} a_j \Pi^\rho_{\vb_j} \]
is the projection operator onto the bi-invariant subspace
corresponding to $\rho$.  Then
\[ P_j = P(\rho) P(\rho,\vb_j) \]
where
\begin{eqnarray}
P(\rho) & = & \norm{\Pi^\rho \ket{H}}^2 \label{eq:prho} \\
P(\rho,\vb_j) & = & a_j \left. \norm{\Pi^\rho_{\vb_j} \ket{Hg}}^2 \right/ P(\rho) 
\enspace . \label{eq:prhoi}
\end{eqnarray}
Note that $P(\rho,\vb_j)$ depends on $g$ but $P(\rho)$ does not, which is
why weak sampling is incapable of distinguishing conjugate subgroups.

\subsection{The probability distribution for a conjugate subgroup}

Now let us use the fact that $\ket{H}$ is a superposition over a
subgroup, and calculate $P(\rho)$ and $P(\rho,\vb_j)$ as defined
in~\eqref{eq:prho} and~\eqref{eq:prhoi}.  This will set the stage for
asking whether we can distinguish different conjugates of $H$ from
each other or even the trivial subgroup.

Fix an irreducible representation $\rho$ that acts on a vector space
$V$.  Then Fourier transforming the state
\[ \ket{H} = \frac{1}{\sqrt{|H|}} \sum_{h \in H} \ket{h} \]
yields the coefficient
\[ \widehat{H}(\rho) = \sqrt{\frac{d_\rho}{|H| |G|}} \sum_{h \in H} \rho(h) 
= \sqrt{\frac{d_\rho |H|}{|G|}} \,\Pi_H \] where $\Pi_H = (1/|H|) \sum_{h \in H} \rho(h)$ is a projection
operator onto a subspace of $V$.  The probability that we observe $\rho$
is then the norm squared of this coefficient,
\begin{equation}
\label{eq:prho2}
 P(\rho) = \norm{\widehat{H}(\rho)}^2 = \frac{d_\rho |H|}{|G|} \,\rank \Pi_H 
\end{equation}
and, as stated above, this is the same for all conjugates $H^g$.  The
conditional probability that we observe the vector $\vb_j$, given that
we observe $\rho$, is then
\begin{equation}
\label{eq:prhoi2}
 P(\rho,\vb_j) = a_j \frac{\norm{\Pi^\rho_{\vb_j} \ket{H} }^2}{P(\rho)}
= a_j \frac{\norm{\widehat{H}(\rho) \vb_j}^2}{P(\rho)}
= a_j \frac{\norm{\Pi_H \vb_j}^2}{\rank \Pi_H} \enspace . 
\end{equation}
In the case where $H$ is the trivial subgroup, $\Pi_H = \one_{d_\rho}$
and $P(\rho,\vb_j)$ is given by
\begin{equation}
\label{eq:natural}
 P(\rho,\vb_j) = \frac{a_j}{d_\rho} \enspace . 
\end{equation}
We call this the \emph{natural distribution} on the frame $B=\{ \vb_j
\}$.  In the case that $B$ is an orthonormal basis, $a_j = 1$ and this
is simply the uniform distribution.

This probability distribution over $B$ changes for a conjugate $H^g$
in the following way.  Again ignoring left multiplication since the
columns are left-invariant, the Fourier transform becomes
\[ \widehat{Hg}(\rho) = \sqrt{\frac{d_\rho |H|}{|G|}} \,\Pi_H \rho(g) \]
and we have 
\[ P(\rho,\vb_j) = a_j \frac{\norm{\Pi_H g \vb_j}^2}{\rank \Pi_H} \]
where we write $g \vb$ for $\rho(g) \vb$. It is not hard to show that,
for any $\vb \in V$, the \emph{expected} value of $\norm{\Pi_H(g
  \vb)}^2$, over the choice of $g \in G$, is $\rank \Pi_H / d_\rho$. Our
primary technical contribution is a method for establishing
concentration results for this random variable.

\section{The variance of projection through a random involution}
\label{sec:var}

In this section we focus on the case where $H=\{1,m\}$ for an element
$m$ chosen uniformly at random from a conjugacy class $[m]$ of
involutions. (Observe that order is preserved under conjugation so
that if $m$ is an involution, then so are all elements of $[m]$.)
Given an irreducible representation $\rho: G \to \U(V)$ and a vector $\vb
\in V$, we bound the variance, over the choice of $m$, of the
probability $P(\rho,\vb)$ that $\vb$ is observed given that we observed
$\rho$.  Our key insight is that this variance depends on how the tensor
product representation $\rho \otimes \rho^*$ decomposes into irreducible
representations $\sigma$, and how the vector $\vb \otimes \vb^*$ projects into
the constituent orthogonal irreducibles.

Recall that, if a representation $\rho$ is reducible, it can be written
as an orthogonal direct sum of irreducibles $\rho = \bigoplus_{\sigma \prec \rho} a_\sigma
\sigma$ where $a_\sigma$ is the multiplicity of $\sigma$.  We let $\Pi^\rho_\sigma$
denote the projection operator whose image is $a_\sigma \sigma$, that is, the
span of all the irreducible subspaces isomorphic to $\sigma$.

\begin{lemma}
\label{lem:exp}
Let $\rho$ be a representation of a group $G$ acting on a space $V$ and
let $\vb \in V$.  Let $m$ be an element chosen uniformly from a
conjugacy class $[m]$ of involutions.  If $\rho$ is irreducible, then
\[ \Exp_m \langle \vb, m \vb \rangle = \frac{\chi_\rho([m])}{\dim \rho} \norm{\vb}^2 \enspace . \]
If $\rho$ is reducible, then 
\[ \Exp_m \langle \vb, m \vb \rangle 
= \sum_{\sigma \prec \rho} \frac{\chi_\sigma([m])}{\dim \sigma} \norm{\Pi^\rho_\sigma \vb}^2 \enspace . \]
\end{lemma}

\begin{proof}
  Fix a particular element $\mu \in [m]$.  Let $\rho([m])$ denote the average
  of $\rho(m)$ over $[m]$; this is
  \[ 
  \rho([m]) = \frac{1}{|[m]|} \sum_{m \in [m]} \rho(m) = \frac{1}{|G|} \sum_{g \in G} \rho(g^{-1} \mu g) 
  = \frac{1}{|G|} \sum_{g \in G} \rho(g)^\dagger \rho(\mu) \,\rho(g) \enspace . 
  \]
  Observe that $\rho([m])$ commutes with $\rho(g)$ for all $g \in G$ and
  hence, by Schur's lemma, its action on any irreducible subspace is
  multiplication by a scalar. Note that for the scalar linear operator
  $A = \alpha \one_d$ acting on a space of dimension $d$, we have $\alpha = \tr
  A / d$. In particular, if $\rho$ is irreducible then
  \[ \rho([m]) = \frac{\chi_\rho([m])}{\dim \rho} \,\one_{d_\rho} \]
  and so
  \[
  \Exp_m \langle \vb, m \vb \rangle 
  = \langle \vb, \rho([m]) \vb \rangle 
  = \frac{\chi_\rho([m])}{\dim \rho} \norm{\vb}^2 
  \enspace . 
  \]
  If $\rho$ is reducible, these same considerations apply to each
  irreducible subspace:
  \[ \rho([m]) = \sum_{\sigma \prec \rho} \frac{\chi_\sigma([m])}{\dim \sigma} \Pi^\rho_\sigma \]
  and so
  \[
  \Exp_m \langle \vb, m \vb \rangle 
  = \langle \vb, \rho([m]) \vb \rangle 
  = \sum_{\sigma \prec \rho} \frac{\chi_\sigma([m])}{\dim \sigma} \langle \vb, \Pi^\rho_\sigma \vb \rangle 
  =  \sum_{\sigma \prec \rho} \frac{\chi_\sigma([m])}{\dim \sigma} \norm{\Pi^\rho_\sigma \vb}^2
  \enspace . 
  \]
\end{proof}
Turning now to the second moment of $\langle \vb, m\vb\rangle$, we observe that 
$$
|\langle \vb, m \vb\rangle|^2 = \langle \vb, m \vb\rangle\langle \vb, m \vb\rangle^* = \langle \vb \otimes \vb^*, m\vb \otimes m\vb^*\rangle  = \langle \vb \otimes \vb^*, m(\vb \otimes \vb^*)\rangle, 
$$
where the action of $m$ on the vector $\vb \otimes \vb^*$ is precisely
given by the action of $G$ in the representations $\rho \otimes \rho^*$. This
will allow us to express the second moment of the inner product $\langle
\vb, m \vb \rangle$ in terms of the projections of $\vb \otimes \vb^*$ into the
irreducible constituents of the tensor product representation $\rho \otimes
\rho^*$.
\begin{lemma}
\label{lem:second}
Let $\rho$ be a representation of a group $G$ acting on a space $V$ and
let $\vb \in V$.  Let $m$ be an element chosen uniformly at random from a
conjugacy class $[m]$ of involutions.  Then
\[ \Exp_m \abs{\langle \vb, m \vb \rangle}^2 
= \sum_{\sigma \prec \rho \otimes \rho^*} 
\frac{\chi_\sigma([m])}{\dim \sigma} \norm{\Pi^{\rho \otimes \rho^*}_\sigma (\vb \otimes \vb^*)}^2 
\enspace . \]
\end{lemma}

\begin{proof}  We write the second moment as a first moment over the product representation $\rho \otimes \rho^*$: as above, $\abs{\langle \vb, m \vb \rangle}^2 = \langle \vb \otimes \vb^*, m(\vb \otimes \vb^*) \rangle$, so that
\[ \Exp_m \abs{\langle \vb, m \vb \rangle}^2 
= \Exp_m \langle \vb \otimes \vb^*, m(\vb \otimes \vb^*) \rangle
\]
and applying Lemma~\ref{lem:exp} completes the proof.
\end{proof}

Now let $\Pi_m = \Pi_H$ denote the projection operator given by
$$
\Pi_m \vv = \frac{\vv + m \vv}{2}\enspace.
$$
For a given vector $\vb \in B$, we will focus on the expectation and variance of $\norm{\Pi_m \vb}^2$.  These are given by the following lemma.

\begin{lemma}  
\label{lem:var}
Let $\rho$ be an irreducible representation acting on a space $V$ and
let $\vb \in V$.  Let $m$ be an element chosen uniformly at random from a
conjugacy class $[m]$ of involutions.  Then
\begin{eqnarray}
  \Exp_{m} \norm{\Pi_m \vb}^2
  & = & \frac{1}{2} \norm{\vb}^2 \left(1 + \frac{\chi_\rho([m])}{\dim \rho} \right) 
  \label{eq:exp} \\
  \Var_{m} \norm{\Pi_m \vb}^2 
  & \leq & \frac{1}{4} 
 \sum_{\sigma \prec \rho \otimes \rho^*} \frac{\chi_\sigma([m])}{\dim \sigma}
   \norm{\Pi^{\rho \otimes \rho^*}_\sigma(\vb \otimes \vb^*)}^2 
  \enspace.
  \label{eq:var} 
\end{eqnarray}
\end{lemma}

\begin{proof}
For the expectation,
\begin{eqnarray*}
\Exp_m \norm{\Pi_m \vb}^2 
& = & \Exp_m \langle \vb, \Pi_m \vb \rangle \\
& = & \frac{1}{2} \Exp_m \left( \langle \vb, \vb \rangle 
+ \langle \vb, m \vb \rangle \right) \\
& = & \frac{1}{2} \norm{\vb}^2 \left(1 + \frac{\chi_\rho([m])}{\dim \rho} \right) 
\end{eqnarray*}
where the last equality follows from Lemma~\ref{lem:exp}.  

For the variance, we first calculate the second moment,
\begin{eqnarray*}
\Exp_m \norm{\Pi_m \vb}^4 
& = & \Exp_m \abs{\langle \vb, \Pi_m \vb \rangle}^2 \\
& = & \frac{1}{4} \Exp_m \abs{\langle \vb, \vb \rangle 
+ \langle \vb, m \vb \rangle}^2 \\
& = & \frac{1}{4} \Exp_m \left( \abs{\langle \vb, \vb \rangle}^2 + 
2 \Re \langle \vb, \vb \rangle \langle \vb, m \vb \rangle 
+ \abs{\langle \vb, m \vb \rangle}^2 \right) \\
& = & \frac{1}{4} \left( \norm{\vb}^4 + 2 \norm{\vb}^4 \frac{\chi_\rho([m])}{\dim \rho} 
+  \sum_{\sigma \prec \rho \otimes \rho^*} \frac{\chi_\sigma([m])}{\dim \sigma}
   \norm{\Pi^{\rho \otimes \rho^*}_\sigma(\vb \otimes \vb^*)}^2 
\right)
\end{eqnarray*}
where in the last line we applied Lemmas~\ref{lem:exp}
and~\ref{lem:second} and the fact that any character evaluated at an
involution is real.  Then
\begin{eqnarray}
\Var_{m} \norm{\Pi_m \vb}^2 
  & = & \Exp_m \norm{\Pi_m \vb}^4 - \left( \Exp_{m} \norm{\Pi_m \vb}^2 \right)^2 \nonumber \\
  & = & \frac{1}{4} \left[ \sum_{\sigma \prec \rho \otimes \rho^*} \frac{\chi_\rho([m])}{\dim \rho}
  \norm{\Pi^{\rho \otimes \rho^*}_\sigma (\vb \otimes \vb^*)}^2 - \norm{\vb}^4
  \left(\frac{\chi_\rho([m])}{\dim \rho} \right)^{\!2\,} \right] \label{eq:var1}
  \enspace .
\end{eqnarray}
Ignoring the second term, which is negative, gives the stated result.
\end{proof}

Finally, we point out that since
\[ 
\Exp_{m} \norm{\Pi_m \vb}^2 = \norm{\vb}^2 \frac{\rank \Pi_m}{\dim \rho}
\]
we have
\begin{equation}
\label{eq:rank}
\frac{\rank \Pi_m}{\dim \rho} = \frac{1}{2}  \left(1 + \frac{\chi_\rho([m])}{\dim \rho} \right)
\enspace ,
\end{equation}
a fact which we will use below.

\section{The representation theory of the symmetric group}

In this section we record the particular properties of $S_n$ and its
representation theory applied in the proofs of the main results.  The
irreducible representations of $S_n$ are labeled by Young diagrams, or
equivalently by integer partitions of $n$,
\[ \lambda = (\lambda_1, \ldots, \lambda_t) \]
where $\sum_i \lambda_i = n$ and $\lambda_i \geq \lambda_{i+1}$ for all $i$.  The
\emph{conjugate} Young diagram $\lambda'$ is obtained by flipping $\lambda$
about the diagonal: $\lambda' = (\lambda'_1, \ldots, \lambda'_{\lambda_1})$ where $\lambda'_j = |\{
i \mid \lambda_i \geq j\}|$. In particular, $\lambda'_1 = t$.  The number of such
diagrams, equal to the number of conjugacy classes in $S_n$, is the
partition number $p(n)$, which obeys
\[ p(n) = (1+o(1)) \frac{1}{4\sqrt{3}\cdot n} \,e^{\pi \sqrt{2n/3}} = e^{\Theta(\sqrt{n})} \enspace . \]
We denote these irreducibles $S^\lambda$, their characters $\chi^\lambda$, and
their dimensions $d^\lambda$.  In particular, $S^{\lambda}$ is the trivial or
parity representation if $\lambda$ is a single row $(n)$ or a single column
$(1,\ldots,1)$ respectively, and $S^{\lambda'}$ is the (tensor) product of
$S^\lambda$ with the parity representation.

The dimensions of the representations of the symmetric group are given
by the remarkable \emph{hook length formula}:
$$
d^\lambda = \frac{n!}{\prod_c \hook(c)}\enspace,
$$
where this product runs over all cells of the Young diagram
associated with $\lambda$ and $\hook(c)$ is the number of cells
appearing in either the same column or row as $c$, excluding those
that are above or to the left of $c$. 

For example, the partition
$\lambda = (\lambda_1, \lambda_2, \lambda_3, \lambda_4) = (6,5,3,2)$
is associated with the diagram shown in Figure~\ref{fig:young}
below. The hook associated with the cell $(2,2)$ in this diagram
appears in Figure~\ref{fig:hook}; it has length 6.

\noindent
\begin{figure}[ht]
  \begin{minipage}[b]{.45\linewidth}
    \begin{center}
      \includegraphics[width=\linewidth]{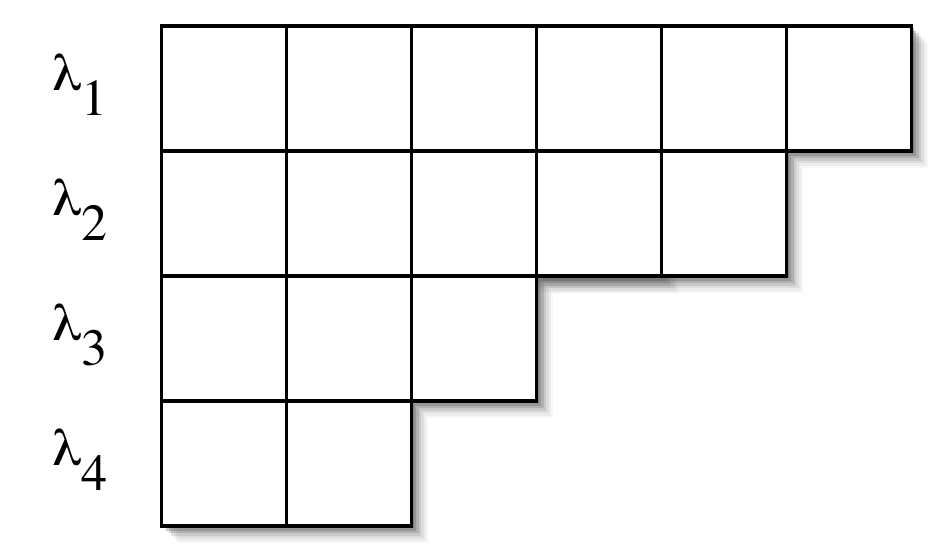}
      \caption{\label{fig:young}The Young diagram for $\lambda = (6,5,3,2)$.}
    \end{center}
  \end{minipage}\hfill
  \begin{minipage}[b]{.45\linewidth}
    \begin{center}
      \includegraphics[width=\linewidth]{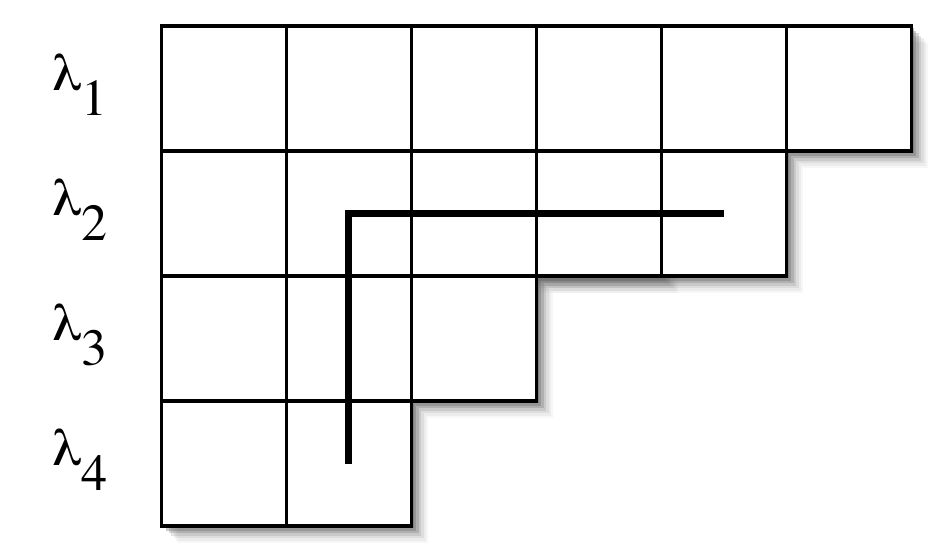}
      \caption{\label{fig:hook}A hook of length 6.}
    \end{center}
  \end{minipage}
\end{figure}

The symmetric groups have the property that every representation
$S^\lambda$ possesses a basis in which its matrix elements are real, and so
all its characters are real.  However, in a given basis $S^\lambda$ might
be complex, so we will refer below to the complex conjugate
representation $(S^\lambda)^*$ (not to be confused with $S^{\lambda'}$).

The study of the asymptotic properties of the representations of $S_n$
typically focuses on the \emph{Plancherel} distribution. (See, e.g.,
Kerov's monograph~\cite{Kerov} for further discussion.)  For a general
group $G$, this is the probability distribution obtained on
$\widehat{G}$ by assigning $\rho$ the probability density $d_\rho^2 /
|G|$. One advantage of this distribution is that the density at $\rho$
is proportional to its contribution, dimensionwise, to the group
algebra $\CG$; this will allow us to reduce structural questions about
representations chosen according to this distribution to questions
about the regular and conjugation representation. Note that in the
context of the hidden subgroup problem, the Plancherel distribution is
exactly the one obtained by performing weak Fourier sampling on the
trivial hidden subgroup.

In the symmetric groups a fair amount is known about representations
chosen according to the Plancherel distribution.  In particular,
Vershik and Kerov have shown that with high probability they have
dimension equal to $e^{\Theta(\sqrt{n})} \sqrt{n!}$.
\begin{theorem}[\cite{VershikK}]
  \label{thm:plancherel}
  Let $S^\lambda$ be chosen from $\widehat{S_n}$ according to the Plancherel
  distribution. Then there exist positive constants $c_1$ and $c_2$
  for which
  $$
  \lim_{n \to \infty} \Pr\left[ e^{-c_1 \sqrt{n}} \sqrt{n!} \leq d^\lambda \leq
    e^{-c_2 \sqrt{n}} \sqrt{n!} \right] = 1\enspace.
  $$
\end{theorem}
Vershik and Kerov have also obtained estimates for the \emph{maximum}
dimension of a representation in $\widehat{S_n}$:
\begin{theorem}[\cite{VershikK}]\label{thm:max-dimension}
  There exist positive constants $\check{c}$ and $\hat{c}$ such that for all $n \geq 1$,
  $$
  e^{-\check{c} \sqrt{n}} \sqrt{n!} \leq 
  \max_{S^\lambda \in \widehat{S_n}} d^\lambda 
  \leq e^{-\hat{c}\sqrt{n}} \sqrt{n!} \enspace .
  $$
\end{theorem}
Along with these estimates, we shall require some (one-sided)
large-deviation versions of Theorem~\ref{thm:plancherel}, recorded below.
\begin{lemma}
  \label{lem:lower}
  Let $S^\lambda$ be chosen according to the Plancherel distribution on
  $\widehat{S_n}$.
  \begin{enumerate}
  \item\label{item:lower-sqrt} Let $\delta = \pi \sqrt{2/3}$. Then for
    sufficiently large $n$, $\Pr\left[ d^\lambda \leq e^{-\delta \sqrt{n}}
      \sqrt{n!} \right] < e^{-\delta\sqrt{n}}$.
  \item\label{item:lower-linear} Let $0 < c < 1/2$. Then $\Pr[ d^\lambda \leq
    n^{cn} ] = n^{-\Omega(n)}$.
  \end{enumerate}
\end{lemma}

\begin{proof} For the first bound, setting $d = e^{-\delta \sqrt{n}}
  \sqrt{n!}$ and using $p(n) < e^{\delta\sqrt{n}}$, we have
  $$
  \sum_{S^\lambda: d^\lambda \leq d} \frac{(d^\lambda)^2}{n!} \leq p(n) \frac{d^2}{n!}  <
  e^{-\delta\sqrt{n}}\enspace.
  $$
  For the second bound, recalling Sterling's approximation $n! \sim
  \sqrt{2 \pi n} (n/e)^n$, we have
    $$
    \frac{1}{n!} \sum_{\lambda: d^\lambda \leq n^{cn}} (d^\mu)^2 \leq \frac{p(cn)
      n^{2cn}}{n!} = n^{-(1 - 2c) n} e^{O(n)} = n^{-\Omega(n)} \enspace .
    $$
\end{proof}

Finally, we will also apply Roichman's~\cite{Roichman96} estimates for
the characters of the symmetric group:
\begin{definition}
  For a permutation $\pi \in S_n$, define the \emph{support} of $\pi$,
  denoted $\supp(\pi)$, to be the cardinality of the set $\{ k \in [n] \mid
  \pi(k) \neq k \}$.
\end{definition}

\begin{theorem}[\cite{Roichman96}] \label{thm:roichman}
  There exist constants $b > 0$ and $0 < q < 1$ so that for $n > 4$,
  for every conjugacy class $C$ of $S_n$, and every irreducible
  representation $S^\lambda$ of $S_n$,
  $$
  \abs{\frac{\chi^\lambda(C)}{d^\lambda}} 
  \leq \left(\max\Bigl(q, \frac{\lambda_1}{n}, \frac{\lambda'_1}{n}\Bigr)\right)^{b \cdot \supp(C)}
  \enspace,
  $$
  where $\supp(C) = \supp(\pi)$ for any $\pi \in C$.
\end{theorem}

In our application, we take $n$ to be even and consider involutions
$m$ in the conjugacy class of elements consisting of $n/2$ disjoint
transpositions, $M = M_n = \{ \sigma \,((12)(34)\cdots(n-1 \enspace n))\, \sigma^{-1} \mid
\sigma \in S_n\}$.  Note that each $m \in M_n$ is associated with one of the
$(n-1)!!$ perfect matchings of $n$ things, and that $\supp(m) = n$.

\section{Strong Fourier sampling over $S_n$}
\label{sec:single}

We consider the hidden subgroup $H= \{1,m\}$, where $m$ is chosen
uniformly from $M = M_n \subset S_{n}$, the conjugacy class
$$
\{ \pi^{-1}((1\; 2)(3\; 4) \cdots (n-1\; n)) \pi \mid \pi \in S_n \}\enspace;
$$
we assume throughout that $n$ is even.  We start by measuring the
name of an irreducible representation, yielding $S^\lambda$ for a diagram
$\lambda$. We remark that Hallgren, Russell, and Ta-Shma~\cite{HallgrenRT00}
established that the probability distribution on $\lambda$ is exponentially
close to the Plancherel distribution in total variation.  We allow
the algorithm designer to choose an arbitrary POVM, with a frame 
$B=\{ \vb_j \}$ and weights $\{ a_j \}$ 
obeying the completeness condition~\eqref{eq:completerho}.  
We will show that with high probability (over $m$ and $\lambda$), 
the conditional distribution induced on the vectors $B$ 
is exponentially close to the natural distribution~\eqref{eq:natural} on $B$.  
It will follow by the triangle inequality that it requires an exponential number of 
single-register experiments to distinguish two involutions from each other or, in
fact, distinguish $H$ from the trivial subgroup.

For simplicity, and to illustrate our techniques, we first prove this for a von Neumann
measurement, i.e., where $B$ is an orthonormal basis for $S^\lambda$.  In this case, 
we show that the probability distribution on $B$ 
(or equivalently, on the columns of $S^\lambda$) 
is exponentially close to the uniform distribution.

\subsection{von Neumann measurements}

\begin{theorem}
\label{thm:one}
Let $B=\{\vb\}$ be an orthonormal basis for an
irreducible representation $S^\lambda$.  
Given the hidden subgroup $H=\{1,m\}$ where $m$ is chosen uniformly
at random from $M$, let $P_m(\vb)$ be the probability
that we observe the vector $\vb$ conditioned on having observed
the representation name $S^\lambda$, and let $U$ be the uniform
distribution on $B$.  Then there is a constant $\delta > 0$ such that for
sufficiently large $n$, with probability at least $1-e^{-\delta n}$ in $m$
and $\lambda$, we have
\[
\norm{ P_m - U }_1 < e^{-\delta n} \enspace .
\]
\end{theorem}

\begin{proof}  First, recall from~\eqref{eq:prhoi2} in Section~\ref{sec:optimal} that the conditional distribution on $B$ is given by (since $a_j = 1$)
\begin{equation}
\label{eq:pb}
P_m(\vb) = P(S^\lambda,\vb) = \frac{\norm{\Pi_m \vb}^2}{\rank \Pi_m} \enspace .
\end{equation}
Our strategy will be to bound $\Var_m \norm{\Pi_m \vb}^2$ using
Lemma~\ref{lem:var}, and apply Chebyshev's inequality to conclude that
it is almost certainly close to its expectation. Recall, however, that
our bounds on the variance of $\norm{\Pi_m \vb}^2$ depend on the
decomposition of $S^\lambda \otimes (S^\lambda)^*$ is into irreducibles and,
furthermore, on the projection of $\vb \otimes \vb^*$ into these
irreducible subspaces. Matters are somewhat complicated by the fact
that certain $S^\mu$ appearing in $S^\lambda \otimes (S^\lambda)^*$ may contribute
more to the variance than others. While Theorem~\ref{thm:roichman}
allows us to bound the contribution of those constituent irreducible
representations $S^\mu$ for which $\mu_1$ and $\mu'_1$ are much smaller
than $n$, those which violate this condition could conceivably
contribute large terms to the variance estimates.  Fortunately, we
will see that the total fraction of the space $S^\lambda \otimes (S^\lambda)^*$,
dimensionwise, consisting of such $S^\mu$ is small with overwhelming
probability. Despite this, we cannot preclude the possibility that for
a \emph{specific} vector $\vb$, the quantity $\Var \norm{\Pi_m \vb}^2$
is large, as $\vb$ may project solely into spaces of the type
described above. On the other hand, as these troublesome spaces amount
to a small fraction of $S^\lambda \otimes (S^\lambda)^*$, only a few $\vb$ can have this property,
and we will see that this suffices to control the distance in total
variation to the uniform distribution.

Specifically, let $0 < c < 1/4$ be a constant, and let $\Lambda = \Lambda_c$
denote the collection of Young diagrams $\mu$ with the property that
either $\mu_1 \geq (1-c) n$ or $\mu_1' \geq (1-c) n$.  We have the following
upper bounds on the cardinality of $\Lambda$ and the dimension of any
$S^\mu$ with $\mu \in \Lambda$:
\begin{lemma}  \label{lem:lambda}
  Let $p(n)$ denote the number of integer partitions of $n$.  Then
  $|\Lambda| \leq 2 cn p(cn)$, and $d^\mu < n^{cn}$ for any $\mu \in \Lambda$.
\end{lemma}

\begin{proof}
  For the first statement, note that removing the top row of a Young
  diagram $\mu$ with $\mu_1 \geq (1-c)n$ gives a Young diagram of size
  $n-\mu_1 \leq cn$.  The number of these is at most $p(cn)$, and summing
  over all such $\mu_1$ gives $cn p(cn)$.  The case $\mu'_1 \geq (1-c)n$
  is similar, and summing the two gives $|\Lambda| \leq 2 cn p(cn)$.
  
  Now let $\mu \in \Lambda$ with $\mu_1 \geq (1-c)n$.  By the hook-length
  formula, since the $i$th cell from the right in the top row has
  $\hook(c) \geq i$, $d^\mu < n!/\mu_1! \leq n!/((1-c)n)! \leq n^{cn}$.  The
  case $\mu'_1 \geq (1-c)n$ is similar.
\end{proof}

As a result, the representations associated with diagrams in $\Lambda$
constitute a negligible fraction of $\widehat{S_n}$; specifically,
from Lemma~\ref{lem:lower}, part~\ref{item:lower-linear}, the
probability that a $\lambda$ drawn according to the Plancherel distribution
falls into $\Lambda$ is $n^{-\Omega(n)}$.
The following lemma shows that this is also true for the distribution
$P(\rho)$ induced on $\widehat{S_n}$ by weak Fourier sampling the coset state
$\ket{H}$.
\begin{lemma}  
\label{lem:notinlambda}
Let $d < 1/2$ be a constant and let $n$ be sufficiently large.  Then
there is a constant $\gamma > 0$ such that
we observe a representation $S^\lambda$ with $d^\lambda \geq 
n^{dn}$ with probability at least $1-n^{-\gamma n}$.
\end{lemma}

\begin{proof}
  The proof is nearly identical to that of Lemma~\ref{lem:lower}, and
  follows an argument from~\cite{GrigniSVV01}.  Recall from~\eqref{eq:prho2} in Section~\ref{sec:optimal} that we
  observe an irreducible $\rho$ with probability $P(\rho) = (d_\rho |H| /
  |G|) \rank \pi_H$.  Given $|H| = 2$, $|G| = n!$, and $\rank \pi_H \leq
  d_\rho$, we have $P(\rho) \leq 2 d_\rho^2 / n!$.  Since the total number of
  irreducibles of $S_n$ is $p(n)$ and $n! > n^n e^{-n}$, we have
  \[ 
  \sum_{S^\lambda: d^\lambda < n^{dn}} P(S^\lambda) < 2 p(n) n^{2dn} / n!  < 2 p(n) e^n
  n^{(2d-1)n}
  \]
  and setting $\gamma < 1-2d$ completes the proof.
\end{proof}

On the other hand, for a representation $S^\mu$ with $\mu \notin \Lambda$,
Theorem~\ref{thm:roichman} implies that
\begin{equation}
  \label{eq:small}
  \abs{\frac{\chi^\mu(M)}{d^\mu}}
  \leq \bigl( \max(q,1-c) \bigr)^{bn} \leq e^{-\alpha n} \enspace
\end{equation}
for a constant $\alpha \geq bc > 0$.  Thus the contribution of such an
irreducible to the variance estimate of Lemma~\ref{lem:var} is
exponentially small.  In addition, let $E_0$ be the event that 
$d^\lambda \geq n^{dn}$ as in Lemma~\ref{lem:notinlambda}; 
then conditioning on $E_0$ and setting $d$ such that $c < 1/4 < d$, 
Lemma~\ref{lem:lambda} implies that $\lambda \notin \Lambda$, 
and Equations~\eqref{eq:rank} and~\eqref{eq:small} imply
\begin{equation}
\label{eq:approx-rank}
 \frac{d^\lambda}{2} \left( 1 - e^{-\alpha n} \right) 
 \leq \rank \Pi_m 
 \leq  \frac{d^\lambda}{2} \left( 1 + e^{-\alpha n} \right)\enspace.
\end{equation}

We turn now to the problem of bounding the multiplicities with which
representations $S^\mu$, for $\mu \in \Lambda$, can appear in $S^\lambda \otimes
(S^\lambda)^*$. While no explicit decomposition is known for $S^\lambda \otimes
(S^\lambda)^*$, the \emph{endomorphism representations} of $S_n$, we
record a coarse bound below which will suffice for our purposes.
Recall that character of $S^\lambda \otimes (S^\lambda)^*$ is $\chi^\lambda \cdot (\chi^\lambda)^* =
(\chi^\lambda)^2$ as characters of $S_n$ are real. The multiplicity of the
representation $S^\mu$ in $S^\lambda \otimes (S^\lambda)^*$ is $\chlangle \chi^\mu,
(\chi^\lambda)^2 \chrangle{G}$. However, this is equal to $\chlangle \chi^\mu
\chi^\lambda, \chi^\lambda \chrangle{G}$, the multiplicity of $S^\lambda$ in the
representation $S^\mu \otimes S^\lambda$.  Counting dimensions, this is clearly
no more than $\dim (S^\mu \otimes S^\lambda) / \dim S^\lambda = d^\mu$. Hence the
multiplicity of $S^\mu$ in $S^\lambda \otimes (S^\lambda)^*$ is never more than
$d^\mu$; we have
\begin{equation}\label{eqn:multiplicity}
  \chlangle \chi^\mu, (\chi^\lambda)^2 \chrangle{G} \leq d^\mu\enspace.
\end{equation}

Let $L \subset S^\lambda \otimes (S^\lambda)^*$ be the subspace consisting of copies of
representations $S^\mu$ with $\mu \in \Lambda$, and let $\Pi_L$ be the
projection operator onto this subspace.  By Lemma~\ref{lem:lambda}, we
have
\[ 
\dim L \leq \sum_{\mu \in \Lambda} (d^\mu)^2 \leq 2 cn p(cn) n^{2cn} = e^{O(\sqrt{n})} n^{2cn}
\enspace .
\]
Note by Lemma~\ref{lem:notinlambda} with $d > 2c$, $\dim L$ is a vanishingly small fraction of $d^\lambda$.

As $B$ is an orthonormal basis for $S^\lambda$, the vectors $\{ \vb \otimes \vb^* \mid \vb \in B\}$ 
are mutually orthogonal in $S^\lambda \otimes (S^\lambda)^*$.  Therefore,
$$
\sum_{\vb \in B} \norm{\Pi_L(\vb \otimes \vb^*)}^2 \leq \dim L \enspace .
$$
In particular, if $B_L \subset B$ denotes the set of $\vb$ with the property that
$$ 
\norm{\Pi_L(\vb \otimes \vb^*)}^2 \geq e^{-\alpha n} \enspace ,
$$
then, conditioning on $E_0$, $|B_L|$ cannot be larger than $e^{\alpha n} \dim L = e^{O(n)} n^{2cn} = n^{-\Omega(n)} d^\lambda$.  For a basis vector $\vb \notin B_L$, we apply Lemma~\ref{lem:var} to bound the variance as follows: assuming pessimistically that $\chi^\mu(M) / d^\mu = 1$ for all $\mu \in \Lambda$, and applying~\eqref{eq:small}, gives
\begin{eqnarray*}
\Var_m \norm{\Pi_m \vb}^2 
& \leq & \frac{1}{4} \left[ 
 \sum_{S^\mu : \mu \in \Lambda} 
   \norm{\Pi^\lambda_\mu(\vb \otimes \vb^*)}^2 
+ \sum_{S^\mu : \mu \notin \Lambda} \frac{\chi^\mu(M)}{d^\mu}
   \norm{\Pi^\lambda_\mu(\vb \otimes \vb^*)}^2 
\right] \\
& \leq & \frac{1}{4} \left[ e^{-\alpha n} + e^{-\alpha n} \sum_{S^\mu : \mu \notin \Lambda} 
\norm{\Pi^\lambda_\mu(\vb \otimes \vb^*)}^2 \right] 
\;\; \leq \;\; \frac{1}{2} \,e^{-\alpha n} \enspace . 
\end{eqnarray*}
In this case, by Chebyshev's inequality, 
\begin{equation} 
\label{eq:chebyshev}
\Pr \left[ \,\abs{ \norm{\Pi_m \vb}^2 - \Exp_m \norm{\Pi_m \vb}^2 }
\geq e^{-\alpha n/3} \right] \leq e^{-\alpha n/3}
\enspace . 
\end{equation}
We say that a basis vector is \emph{bad} if this bound is violated, i.e., 
$$
\abs{ \norm{\Pi_m \vb}^2 - \Exp_m \norm{\Pi_m \vb}^2 }
\geq e^{-\alpha n/3}
\enspace . 
$$
Let $\bad$ denote the subset of $B$ consisting of bad basis vectors
(observe that while $B_L$ depends only on the choice of $\lambda$, $\bad$
depends also on $m$).  Then~\eqref{eq:chebyshev} implies
$$
\Exp_m |\bad| \leq e^{-\alpha n/3} d^\lambda .
$$
Now let $E_1$ be the event that 
\[ |\bad| < e^{-\alpha n/6} d^\lambda \enspace ; \]
then by Markov's inequality, $E_1$ occurs with probability at least $1-e^{-\alpha n/6}$.

Now let us separate $\norm{P_m - U}_1$ into contributions from basis vectors outside and inside
$B_L \cup \bad$:
\begin{equation}
\label{eq:sep}
\norm{P_m - U}_1 
= \sum_{\vb \notin B_L \cup \bad} \abs{P_m(\vb) - U}
+ \sum_{\vb \in B_L \cup \bad} \abs{P_m(\vb) - U}
\enspace .
\end{equation}
The first sum is taken only over vectors $\vb$ for which
\[ \abs{ \norm{ \Pi_m \vb }^2 - \Exp_m \norm{ \Pi_m \vb }^2 }
< e^{-\alpha n/3} \enspace . \]
Conditioning on $E_0$ and recalling that $P_m(\vb) = \norm{\Pi_m \vb}^2/ \rank \!\Pi_m$, the rank estimate of~\eqref{eq:approx-rank} gives
\begin{equation}
\label{eq:out}
 \sum_{\vb \notin B_L \cup \bad} \abs{P_m(\vb) - U}
\leq \frac{e^{-\alpha n/3}}{\rank \Pi_m} \cdot d^\lambda 
\leq \frac{2 e^{-\alpha n/3}}{1-e^{-\alpha n}}
< 4 e^{-\alpha n/3} \enspace . 
\end{equation}
It follows that $P_m(B_L \cup \bad)$ is at most $|B_L \cup \bad|/d^\lambda + 4 e^{-\alpha n/3}$.  
Therefore, since conditioning on $E_0$ and $E_1$ we have $|B_L \cup \bad| \leq (n^{-\Omega(n)} + e^{-\alpha n/6}) d^\lambda < 2e^{-\alpha n/6} d^\lambda$, the second sum in~\eqref{eq:sep} is at most
\begin{equation}
\label{eq:in}
\sum_{\vb \in B_L \cup \bad} \abs{P_m(\vb) - \frac{1}{d^\lambda}} 
\leq P_m(B_L \cup \bad) + \frac{|B_L \cup \bad|}{d^\lambda} 
< 4 e^{-\alpha n/6} + 4 e^{-\alpha n/3} < 5 e^{-\alpha n/6} 
\enspace . 
\end{equation}
Then combining~\eqref{eq:sep},~\eqref{eq:out} and~\eqref{eq:in}, 
\[  \norm{ P_m - U }_1 < 4 e^{-\alpha n/3} + 5 e^{-\alpha n/6} < 6 e^{-\alpha n/6} \]
with probability at least $\Pr[E_0 \land E_1]  \geq 1 - n^{-\gamma n} - e^{-\alpha n/6} \geq 1 - 2e^{-\alpha n/6}$.  We complete the proof by setting $\delta < \alpha/6$.
\end{proof}

\subsection{Arbitrary POVMs}
\label{sec:single-frame}

We now generalize the proof of Theorem~\ref{thm:one} to the case where
the algorithm designer is allowed to choose an arbitrary finite frame $B=\{ \vb \}$ 
of unit length vectors in $S^\lambda$, with a family of positive real
weights $a_\vb$, that satisfy the completeness condition
\begin{equation}
\label{eq:complete2} 
\sum_\vb a_\vb \ket{\vb} \bra{\vb} = \one
\enspace .
\end{equation}
(Note that this is simply~\eqref{eq:completerho} where we have
written $\vb$ and $a_\vb$ instead of $\vb_j$ and $a_j$.) 

\begin{theorem}
\label{thm:one-frames}
Let $B=\{\vb\}$ be a frame with 
weights $\{a_\vb\}$ satisfying the completeness condition~\eqref{eq:complete2} for an
irreducible representation $S^\lambda$.  
Given the hidden subgroup $H=\{1,m\}$ where $m$ is chosen uniformly
at random from $M$, let $P_m(\vb)$ be the probability
that we observe the vector $\vb$ conditioned on having observed
the representation name $S^\lambda$, and let $N$ be the natural 
distribution~\eqref{eq:natural} on $B$.  Then there is a constant $\delta > 0$ such that 
for sufficiently large $n$, with probability at least $1-e^{-\delta n}$ in $m$
and $\lambda$, we have
\[
\norm{ P_m - N }_1 < e^{-\delta n} \enspace .
\]
\end{theorem}

\begin{proof}  Recall from~\eqref{eq:prhoi2} in Section~\ref{sec:optimal} that the conditional distribution on $B$ is given by
\[ P_m(\vb) = P(S^\lambda,\vb) = a_\vb \frac{\norm{\Pi_m \vb}^2}{\rank \Pi_m} \]
and the natural distribution~\eqref{eq:natural} is given by $N(\vb) = a_\vb / d^\lambda$.

The proof of Theorem~\ref{thm:one} goes through with a few
modifications.  First, let us change some semantics: given a subset $A
\subseteq B$, we let $|A|$ denote the weighted size of $A$,
\[ |A| = \sum_{\vb \in A} a_\vb\enspace. \]
With this definition, the total probability that falls in $A$ under the natural distribution is $N(A) = |A|/d^\lambda$.  Then we will use the following lemma:
\begin{lemma} 
\label{lem:framesum} 
Let $L$ be a subspace of $S^\lambda \otimes (S^\lambda)^*$ and let $\Pi_L$ be the projection operator onto $L$.  Then
\begin{equation}
\label{eq:framesum}
\sum_{\vb \in B} a_\vb \norm{ \Pi_L (\vb \otimes \vb^*) }^2 \leq \dim L \enspace .
\end{equation}
\end{lemma}

\begin{proof}
First note that a vector $\ve \in S^\lambda \otimes (S^\lambda)^*$ has entries $\ve_{j,k}$ for $1 \leq j,k \leq d^\lambda$.  There is a unique linear operator $E$ on $S^\lambda$ whose matrix entries are $E_{j,k} = \ve_{j,k}$, and the inner product $\langle \vb \otimes \vb^* , \ve \rangle$ in $S^\lambda \otimes (S^\lambda)^*$ can then be written as the bilinear form $\langle \vb, E \vb \rangle$ in $S^\lambda$.  The Frobenius norm of $E$ is $\norm{E}^2 = \tr E^\dagger E = \norm{\ve}^2$.  
  
Now let $\{ \ve_i \}$ be an orthonormal basis for $L$ and let $E_i$ 
be the operator corresponding to $\ve_i$.  Then 
    \begin{align*}  
    \sum_{\vb \in B} a_\vb \abs{\langle \vb \otimes \vb^*, \ve_i \rangle}^2 &= \sum_{\vb
      \in B} a_\vb \abs{\langle \vb, E_i \vb \rangle}^2 \leq \sum_{\vb \in B} a_\vb
    \norm{\vb}^2 \norm{E_i \vb}^2
    = \sum_{\vb \in B} a_\vb \norm{E_i \vb}^2 \\
    &= \sum_{\vb \in B} a_\vb \,\tr \!\left( E_i^\dagger \ket{\vb} \bra{\vb} E_i \right)
= \tr \!\left[ E_i^\dagger \left( \sum_{\vb \in B} a_\vb \ket{\vb} \bra{\vb} \right) E_i \right] 
= \tr E_i^\dagger E_i = \norm{\ve_i}^2 = 1 
  \end{align*}
where we used the Cauchy-Schwartz inequality in the second line and completeness in the second.
Summing over the $\dim L$ basis vectors $\ve_i$ then gives~\eqref{eq:framesum}.
\end{proof}

We define $\Lambda$ and $E_0$ as before, and Lemmas~\ref{lem:lambda} and~\ref{lem:notinlambda} still apply.  As before, let $L \subset S^\lambda \otimes (S^\lambda)^*$ be the subspace consisting of copies of representations $S^\mu$ with $\mu \in \Lambda$, and let $B_L \subset B$ denote the set of $\vb$ with the property that
$$ 
\norm{\Pi_L(\vb \otimes \vb^*)}^2 \geq e^{-\alpha n} \enspace .
$$
Then Lemma~\ref{lem:framesum} implies that 
\[ |B_L| \leq e^{\alpha n} \dim L \]
where $|B_L|$ is defined as above.  
We again define $\bad$ as the set of $\vb \in B \setminus B_L$ such that
$$
\abs{ \norm{\Pi_m \vb}^2 - \Exp_m \norm{\Pi_m \vb}^2 }
\geq e^{-\alpha n/3}
$$
and Chebyshev's and Markov's inequalities imply that the event $E_1$, namely 
\[ |\bad| < e^{-\alpha n/6} d^\lambda \enspace , \]
occurs with probability at least $1-e^{-\alpha n/6}$.  

We separate $\norm{P_m - N}_1$ as we did $\norm{P_m - U}_1$ before:
\begin{equation}
\label{eq:sep-frame}
\norm{P_m - N}_1 
= \sum_{\vb \notin B_L \cup \bad} \abs{P_m(\vb) - N(\vb)}
+ \sum_{\vb \in B_L \cup \bad} \abs{P_m(\vb) - N(\vb)}
\enspace .
\end{equation}
Since $\abs{ \norm{ \Pi_m \vb }^2 - \Exp_m \norm{ \Pi_m \vb }^2 }
< e^{-\alpha n/3}$ for all $\vb \notin B_L \cup \bad$, and since $\sum_\vb a_\vb = d^\lambda$, 
conditioning on $E_0$ and using~\eqref{eq:approx-rank} bounds the first sum as follows,
\begin{equation}
\label{eq:out-frame}
 \sum_{\vb \notin B_L \cup \bad} \abs{P_m(\vb) - N(\vb)}
\leq \frac{e^{-\alpha n/3}}{\rank \Pi_m} \cdot d^\lambda 
\leq \frac{2 e^{-\alpha n/3}}{1-e^{-\alpha n}}
< 4 e^{-\alpha n/3} \enspace . 
\end{equation}
It follows that $P_m(B_L \cup \bad)$ is at most $|B_L \cup \bad|/d^\lambda + 4
e^{-\alpha n/3}$.  Conditioning on $E_0$ and $E_1$, we have $|B_L \cup \bad| \leq
(n^{-\Omega(n)} + e^{-\alpha n/6}) d^\lambda < 2e^{-\alpha n/6} d^\lambda$.  Thus the second sum
in~\eqref{eq:sep-frame} is at most
\begin{equation}
\label{eq:in-frame}
\sum_{\vb \in B_L \cup \bad} \abs{P_m(\vb) - N(\vb)} 
\leq P_m(B_L \cup \bad) + N(B_L \cup \bad) 
< 4 e^{-\alpha n/6} + 4 e^{-\alpha n/3} < 5 e^{-\alpha n/6} 
\enspace . 
\end{equation}
Then combining~\eqref{eq:sep-frame},~\eqref{eq:out-frame} and~\eqref{eq:in-frame}, 
\[  \norm{ P_m - N }_1 < 4 e^{-\alpha n/3} + 5 e^{-\alpha n/6} < 6 e^{-\alpha n/6} \]
with probability at least $\Pr[E_0 \land E_1]  \geq 1 - n^{-\gamma n} - e^{-\alpha n/6} \geq 1 - 2e^{-\alpha n/6}$.  We complete the proof by setting $\delta < \alpha/6$ as before.
\end{proof}

\section{Structured involutions and the case of Graph Isomorphism}

The preceding development focuses on the case where the hidden
subgroup is distributed uniformly among the conjugates of the subgroup
$H = \{ 1, m \}$. As such, this shows that the canonical reduction of
\textsc{Graph Automorphism} (the problem of determining whether a
given graph has a non-trivial automorphism) to the hidden subgroup
problem does not, in general, give rise to an efficient quantum
algorithm via Fourier sampling. 

However, the canonical reduction of
\textsc{Graph Isomorphism} to the hidden subgroup problem induces a
more structured set of involutions.  As referred to in the Introduction, 
fixing two rigid graphs $G_1 = (V_1, E_1)$ and $G_2 = (V_2, E_2)$, each with $n$
vertices, the automorphism group of their disjoint union $(V_1 \cup V_2, E_1 \cup
E_2)$ is nontrivial exactly when they are isomorphic, in which case it
is generated by an involution $m$ with full support such that $m(V_1) = V_2$. 
Identifying $V_1$ and $V_2$ with the sets $\{1, \ldots, n\}$ and
$\{n+1, \ldots, 2n\}$ respectively, and letting $s$ denote the involution 
$(1\;n+1) (2\; n+2) \ldots (n-1\; 2n)$, the standard reduction to the
hidden subgroup problem in $S_{2n}$ then results in a hidden subgroup 
$H = \{ 1,m \}$ where $m$ is a conjugate involution $a^{-1} s a$.  However, 
rather than $a$ being drawn from all of $S_{2n}$, it is 
an element of the Young subgroup $S_{n,n}$ which fixes $V_1$ and $V_2$:
\remove{
\begin{equation}
  \label{eq:young-conjugation}
  m = a^{-1} s a\enspace,
\end{equation}
}
$$
S_{n,n} = \bigl\{ \pi \in S_{2n} \mid \pi(\{1, \ldots,n\}) = \{1,
\ldots, n\} \bigr\} \cong S_n \times S_n \enspace .
$$
In other words, rather than considering all conjugates of $m$ in
$S_{2n}$, it suffices just to consider conjugates in $S_{n,n}$.  {\em
  A priori}, it seems that this smaller set of possible hidden
subgroups might be easier to identify.  Moreover, let $K$ be the
subgroup generated by $S_{n,n}$ and $s$: this is the wreath product
$S_n \wr \Z_2$, which can also be written as a semidirect product $K =
(S_n \times S_n) \ltimes \Z_2$.  Then each such $H$ is contained in $K$,
and it seems that it might be more intelligent to Fourier sample over
$K$ rather than over all of $S_{2n}$.

However, we can show that nothing is gained by this approach.  First, note that 
the involutions described above form the ($K$-)conjugacy class 
$$
\{ \bigl( (\alpha, \alpha^{-1}), 1 \bigr) \in (S_n \times S_n) \ltimes \Z_2 
\mid \alpha \in S_n \}\enspace.
$$
We remark the that the development of Section~\ref{sec:optimal} is
unchanged, and that the optimal measurement to find a hidden conjugate 
again consists of strong Fourier sampling.

Now note that Fourier sampling over $S_{2n}$ and over $K$ are
equivalent for the following reason: suppose we are trying to
distinguish a set of hidden subgroups $H_i \subset G$, all of which are
contained in a subgroup $K \subset G$.  Let $T$ be a set of representatives
for the cosets of $K$.  Then a random left coset of $H_i$ in $G$ is
the product of a random left coset of $H_i$ in $K$ with a random
element of $T$.  Thus the mixed state describing a uniformly random
coset of $H_i$ in $G$ can be written as the corresponding state over
$K$ with the completely mixed state over $T$.  Since this completely
mixed state (whose density matrix is the identity) contains no
information, nothing is gained (or lost) by sampling over all of $G$
rather than over $K$.

Finally, we observe that one can obtain easy character bounds for this
conjugacy class of $K$.  One determines $K$'s irreducible
representations and their characters as follows.  For two irreducible
representations $\rho$ and $\sigma$ of $S_n$, let $\rho \boxtimes \sigma$ denote their
tensor product as a representation of $S_{n,n} \cong S_n \times S_n$.  We
consider the {\em induced representation} $\Ind_{S_{n,n}}^K (\rho \boxtimes
\sigma)$ (see~\cite{Serre77}) and denote its character $\chi_{\{\rho,\sigma\}}$.  It is
easy to see that
$$
\chi_{\{\rho,\sigma\}}\bigl( ((\alpha, \beta),t) \bigr) = \begin{cases} 0 & \text{if}\;t=1\\
  \chi_\rho(\alpha)\chi_\sigma(\beta) +
  \chi_\sigma(\alpha)\chi_\rho(\beta) & \text{if}\;t=0\enspace;
\end{cases}
$$
as the notation suggests, this depends only on the multiset $\{ \rho, \sigma\}$.
Observe that if $A$ and $B$ are two distinct multisets of
representations of $S_n$ (each of cardinality two) then the characters
$\chi_A$ and $\chi_B$ are orthogonal.  Furthermore, an easy computation
shows that $\langle \chi_{\{\rho,\sigma\}}, \chi_{\{\rho, \sigma\}}\rangle = 1+\delta_{\rho,\sigma}$ and hence that $\chi_{
  \{ \rho,\sigma \} }$ is irreducible if $\rho \not\cong \sigma$, while if $\rho \cong \sigma$ then the
induced representation is the direct sum of two irreducible
representations,
\[ \Ind_{S_{n,n}}^K (\rho \boxtimes \rho)  
\cong ( \rho \boxtimes \rho \otimes \one ) \oplus ( \rho \boxtimes \rho \otimes \pi )
\]
where $\one$ and $\pi$ are the trivial and sign representations, respectively, of $\Z_2$.
Each of these irreducible representations acts on $V_\rho \otimes
V_\rho$, the vector space supporting the action of $\rho \boxtimes \rho$.  
Both of them realize the element $((\alpha,\beta),0)$ as the
linear map $\rho(\alpha) \otimes \rho(\beta)$, while $\rho \boxtimes \rho \otimes \one$ 
and $\rho \boxtimes \rho \otimes \pi$ realize the element %$((\mathbf{1}, \mathbf{1}),1)$ 
$((1,1),1)$ as the maps which send $\vec{u} \otimes \vec{v}$ to $\vec{v} \otimes \vec{u}$ 
and $-\vec{v} \otimes \vec{u}$ respectively. The characters of these
representations can be determined by inspection; in particular, 
$$
\chi_{\{\rho,\rho\},\one} ((\alpha, \beta),t) = \begin{cases} \chi_\rho(\alpha) + \chi_\rho(\beta) & \text{if}\;t = 0\enspace,\\
  \chi_\rho(\alpha\beta) & \text{if}\;t=1\enspace, \end{cases}
\qquad
\chi_{\{\rho,\rho\},\pi} ((\alpha, \beta),t) = \begin{cases} \chi_\rho(\alpha) + \chi_\rho(\beta) & \text{if}\;t = 0\enspace,\\
  -\chi_\rho(\alpha\beta) & \text{if}\;t=1\enspace, \end{cases}
$$
where $\chi_{\{\rho,\rho\},\one}$ and $\chi_{\{\rho,\rho\},\pi}$ denote 
the characters of $\rho \boxtimes \rho \otimes \one$ and $\rho \boxtimes \rho \otimes \pi$ 
respectively.

In light of this description of the irreducible characters of $K$, one
finds that $\chi_{\sigma,\rho}((\alpha, \alpha^{-1}), 1) = 0$ when $\sigma \not\cong \rho$ and
that $|\chi_{\rho \boxtimes \rho, \lambda}((\alpha, \alpha^{-1}), 1))| = d_\rho$ for $\lambda \in
\{\one, \pi\}$. In particular, the absolute value of the normalized
character at $\rho \boxtimes \rho \otimes \lambda$ is $1/d_\rho$. As before, the dimensions
$d_\rho$ of the diagrams with large normalized character sum to very
little, and the analysis of Section~\ref{sec:single} can be undertaken
\emph{mutatis mutandis}. Remarkably, this can be carried out without
use of Roichman's estimates!

\section*{Acknowledgments.}  
This work was supported by NSF grants CCR-0093065, PHY-0200909,
EIA-0218443, EIA-0218563, CCR-0220070, and CCR-0220264, and ARO grant
W911NF-04-R-0009.  We are grateful to Denis Th{\'e}rien, McGill
University, and Bellairs Research Institute for organizing a workshop
at which this work began; to Dorit Aharonov, Daniel Rockmore, and
Umesh Vazirani for helpful conversations; to Chris Lomont for pointing
out several typos; and to Tracy Conrad and Sally Milius for their
support and tolerance.  C.M.\ also thanks Rosemary Moore for providing
a larger perspective. Finally, we thank Gorjan Alagic for his comments
on the structured involutions material.

\end{document}